\documentclass[aps,prd,twocolumn,showpacs,superscriptaddress]{revtex4} 

\usepackage{graphicx}
\usepackage{amsmath}

\IfFileExists{srcltx.sty}{\usepackage[active]{srcltx}}

\newcommand{\ynm}{\ensuremath{Y_{\nu_\mu}}}

\begin{document}

\title{Pulsar Kicks from Active-Sterile Neutrino Transformation in Supernovae}
\author{Chad T. Kishimoto}
\affiliation{Department of Physics and Astronomy, University of California, Los Angeles, California 90095}
\date{\today}

\begin{abstract}
Observations of radio pulsars have revealed that they have large velocities which may be greater than $1000 ~{\rm km}/{\rm s}$.  
In this work, the efficacy of an active-sterile neutrino transformation mechanism to provide these large pulsar kicks is investigated.  A phase-space based approach is adopted to follow the the transformation of active neutrinos to sterile neutrinos through an MSW-like resonance in the protoneutron star to refine an estimate to the magnitude of the pulsar kick that can be generated in such an event.  Estimates of the magnitude of the pulsar kick created and the overall cooling due to sterile neutrino emission suggest that this mechanism can create the large pulsar kicks while not overcooling the star.
%The result is that this mechanism can create the large pulsar kicks that are observed while not overcooling the star.
\end{abstract}
%\pacs{95.35.+d, 14.60.Pq, 14.60.St}
\maketitle

\section{Introduction}
\label{sec:introduction}

Supernova explosions have long been used as a laboratory for particle physics, especially for the existence of light, weakly interacting particles such as axions, majorons, and sterile neutrinos.  In particular, sterile neutrinos are of interest because one with a mass of several keV is a viable -- and possibly observable -- dark matter candidate.  The study of the supernova mechanism can provide limits on sterile neutrino parameters from the requirement that the emission of sterile neutrinos from the protoneutron star does not overcool the star before the emission of active neutrinos which are observed in the SN1987A event.  

Astronomical observations show evidence that radio pulsars can have very large velocities \cite{r70, gl72, c86}.  Estimates of their average velocities are of the order of a few hundred km/s, but some pulsars appear to have velocities in excess of $1000~{\rm km}/{\rm s}$ \cite{hp97, cc98, acc02, hllk05}.  

Radio pulsars are a small subset of the billions of neutron stars in the galaxy; each neutron star being born in a supernova, the cataclysmic end to the life of a massive star.  It would be natural to look to the supernova explosion for a mechanism to generate these large observed velocities of pulsars (``pulsar kicks'').  While hydrodynamical asymmetries in the supernova collapse may be able to generate such large velocities, many have failed.  Recent approaches are based on two-dimensional simulations \cite{nbblo10, spjkm04, skjm06, wjm10}, but it is unclear whether, in three dimensions, this mechanism can account for both the magnitude of the kick and the correlation between the spin axis and pulsar velocity that is observed in the data \cite{johnston05, wlh06, nr07, nr08}.

The signature of keV sterile neutrino emission in a supernova is a significant momentum anisotropy, that generates a large pulsar kick \cite{ks97}.  In addition, the observed spin-velocity correlation is a generic result of keV sterile neutrino emission with a large enough magnitude to generate the observed pulsar kicks.  The observed pulsar kicks can be explained by the emission of sterile neutrinos created by active-sterile neutrino transformation within the highly magnetized, dense core of a protoneutron star.  The purpose of this paper is to understand the physics of sterile neutrino creation and subsequent emission in detail.

Sterile neutrinos are an intriguing extension of the Standard Model toward resolving unexplained phenomena in astrophysics \cite{kusenko09}.  The concept of a sterile neutrino, or ``right-handed neutrino,'' as a neutrino that does not participate in the weak interaction was introduced by Bruno Pontecorvo \cite{pontecorvo67}.  Although the sterile neutrino does not directly couple to the strong, weak or electromagnetic interactions, it may couple to the active neutrinos and/or the Higgs boson which may provide a number of avenues for sterile neutrinos to be an integral component in astroparticle physics.  

These right-handed neutrinos are natural extensions of the Standard Model to account for the fact that neutrinos have mass.  The seesaw mechanism can allow the mass of these right-handed neutrinos to be very large \cite{yanagida79, grs79, minkowski77}, but has also recently been shown that can also allow the mass of these neutrinos to be smaller \cite{kty10}.  It is the latter of these possibilities that are regarded as sterile neutrinos that provide exciting opportunities to explain astrophysical phenomena.

This paper will discuss the prospects of a sterile neutrino with a keV mass and vacuum mixing between the sterile and active neutrinos.  There has been significant interest in a keV mass sterile neutrino as a proposed dark matter candidate \cite{dw94, sf99, afp, dh02, asl07, kf08}, possible mechanism for generating pulsar kicks \cite{ks97, fkmp03, kmm08}, strengthening the supernova explosion \cite{fk06, hf06, hf07}, and as a source of ionization in the reionization epoch \cite{bk06}.  In addition, preliminary astrophysical observations suggest the possibility of the existence of a dark matter candidate sterile neutrino with masses of either $5~{\rm keV}$ \cite{lk10} or $17~{\rm keV}$ \cite{ps10}.

In Section \ref{sec:resonances} an active-sterile neutrino transformation mechanism to produce pulsar kicks will be presented.  Section \ref{sec:calculations} discusses the calculation of the magnitude of the pulsar kick generated, while Section \ref{sec:results} applies the proposed mechanism in a mass element of a protoneutron star to estimate the magnitude of the pulsar kick generated by active-sterile neutrino transformation.  Finally, a brief discussion on the mechanism is presented in Section \ref{sec:discussion} and conclusions are presented in Section \ref{sec:conclusions}.

\section{Resonant Sterile Neutrino Production}
\label{sec:resonances}

Let us consider the evolution of neutrino states during the first $\sim 10~{\rm s}$ after the formation of a protoneutron star.  In this discussion, we will consider the simplest of active-sterile neutrino mixing between one active and one sterile state so that the mass and weak eigenstates are related in vacuum by
\begin{eqnarray}
\vert \nu_1 \rangle & = &\cos \theta \vert \nu_\alpha \rangle - \sin \theta \vert \nu_s \rangle \nonumber \\
\vert \nu_2 \rangle & =  &\sin \theta \vert \nu_\alpha \rangle + \cos \theta \vert \nu_s \rangle ,
\label{eq:vacuum-mixing}
\end{eqnarray}
where $\nu_{1,2}$ are the mass states, $\nu_\alpha$ with $\alpha = e,$ $\mu$, or $\tau$ is the active neutrino, $\nu_s$ is the sterile neutrino, and $\theta$ is the vacuum mixing angle between the active and sterile neutrino.  As the neutrinos travel through matter, the relationship in Eq.~(\ref{eq:vacuum-mixing}) is slightly altered as the mass states become instantaneous mass ({\it i.e.}, energy) states and the vacuum mixing angle is replaced by the effective matter mixing angle $\theta_m$.  The effective matter mixing angle can be related to the forward scattering potential of $\nu_\alpha$ in matter, $V_\alpha$, by
\begin{equation}
\sin^2 2 \theta_m = \frac{(\delta m^2) / 2 p)^2 \sin^2 2 \theta}{(\delta m^2 / 2 p)^2 \sin^2 2 \theta + [ (\delta m^2 / 2 p) \cos 2 \theta - V_\alpha ]^2} ,
\end{equation}
for a neutrino state with momentum $p$ \cite{afp}.  For the sterile neutrino parameters that are relevant in this paper ($\sin^2 \theta \ll 1$, $m_2 \gg m_1$), the difference in the squares of the two neutrino masses, $\delta m^2 = m_2^2 - m_1^2$, can be related to the ``sterile neutrino mass,'' $m_s$, by $\delta m^2 \approx m_s^2$.  

In vacuum, since $\sin^2 \theta \ll 1$, sterile neutrinos are most closely associated with $\nu_2$, the heavier mass state.  However, as the forward scattering potential increases, so does the matter mixing angle.  If the forward scattering potential increases to a point that is analogous to the MSW resonance \cite{MS, W},
\begin{equation}
\frac{m_s^2 \cos 2 \theta}{2 p} = V_\alpha ,
\label{eq:msw}
\end{equation}
then $\theta_m = \pi / 4$; the active and sterile neutrino are maximally mixed.  

The dynamics of the neutrino state can be discussed in terms of three length scales:  the neutrino oscillation length, the width of the resonance discussed above, and the neutrino mean free path.  There are two mechanisms in the conversion of active neutrinos to sterile neutrinos:  non-resonant and resonant production of sterile neutrinos.  

Scattering induced, non-resonant production of sterile neutrinos occurs at a rate proportional to the neutrino scattering rate and $\sin^2 2 \theta_m$.  However, if the neutrino oscillation length is short compared to the scattering length, then sterile neutrino production will be suppressed by the quantum Zeno effect.  

Resonant sterile neutrino production occurs at the MSW-like resonances described by Eq.~(\ref{eq:msw}).  This resonant production mechanism is efficient in converting active neutrinos that meet the resonance condition into sterile neutrinos as long as the neutrino oscillation length at resonance is small compared to the resonance width which is small compared to the mean free path.  A large momentum anisotropy from the conversion of active to sterile neutrinos can be produced by this mechanism.

The neutrino oscillation length is maximized at the resonance point.  The neutrino oscillation length at resonance is
\begin{equation}
\ell_{\rm res} = \frac{2 p}{m_s^2 \sin 2 \theta} .
\end{equation}
For the adiabatic evolution of these neutrinos, the neutrino oscillation length at resonance should be small compared to the width of the resonance, 
\begin{equation}
\delta t = \left\vert \frac{1}{V_\alpha} \frac{d V_\alpha}{d x} \right\vert^{-1} \tan 2 \theta ,
\end{equation}
where $d/dx$ is the spatial derivative along the neutrino trajectory.  Finally, to describe the coherent evolution of the neutrino as it traverses the resonance, the width of the resonance should be small compared to the mean free path of the active neutrino,
\begin{equation}
\lambda_{\rm mfp} = \left( \frac{\rho}{m_n} \sigma_{\rm weak} \right)^{-1} ,
\end{equation}
where $\rho$ is the density, $m_n$ is the neutron mass, and $\sigma_{\rm weak} \sim G_F^2 p^2$ is the weak scattering cross section with $G_F$ the Fermi coupling constant \cite{afp}.

If the width of the resonance is small compared to the neutrino mean free path, we can approach the evolution of neutrinos through the MSW resonances as being coherent.  Neglecting any neutrino magnetic moments, the forward scattering potential for active neutrinos in a magnetized, neutral and degenerate background can be written as \cite{nr88, dnp89, ec96, nssv97, als97}
\begin{align}
%V_e & = &- V_0 (\rho) ( 1 - 3 Y_e - 4 Y_{\nu_e} - 2 Y_{\nu_\mu} - 2 Y_{\nu_\tau} ) , \\
V_e & = - V_0 (\rho) ( 1 - 3 Y_e - 4 Y_{\nu_e} - 2 Y_{\nu_\mu} - 2 Y_{\nu_\tau} ) \nonumber \\
 & \qquad + V_1^e (\rho, B, T; Y_e) \cos \theta_\nu \\
V_\beta & = - V_0 (\rho) f_\beta (Y_\ell) + V_1 (\rho, B, T; Y_e) \cos \theta_\nu ,  
%V_\beta & = - V_0 (\rho) f_\beta (Y_\ell) + \frac{e G_F}{\sqrt{2}} \left( \frac{3 n_e}{\pi^4} \right)^{1/3} \hat{\mathbf{k}} \cdot \vec{\mathbf{B}} , 
\end{align}
for $\beta = \mu$ or $\tau$, $\cos \theta_\nu$ is the angle between the neutrino momentum and magnetic field, where
\begin{equation}
f_\beta (Y_\ell)= 1 - Y_e - 2 Y_{\nu_\beta} - 2 Y_{\nu_e} - 2 Y_{\nu_\mu} - 2 Y_{\nu_\tau} ,
\end{equation}
and
\begin{equation}
V_0 (\rho) = \frac{G_F}{\sqrt{2}} \frac{\rho}{m_N} \approx 3.8 \rho_{14}~{\rm eV}  ,
\end{equation}
with $\rho_{14}$ the density in units of $10^{14}~{\rm g}/{\rm cm}^3$.
The $Y_\ell$ are the local lepton asymmetries and are defined as the lepton number divided by the baryon number, 
\begin{equation}
Y_\ell = \frac{n_\ell - n_{\bar\ell}}{n_b} ,
\end{equation}
where $n_\ell$, $n_{\bar\ell}$ and $n_b$ are the number densities of species $\ell$, its anti-particle $\bar\ell$, and baryons, respectively.
For anti-neutrinos, the forward scattering potential is the negative of the forward scattering potential for neutrinos, $\bar{V}_\alpha = - V_\alpha$, and the forward scattering potential for sterile neutrinos is zero, $V_s = 0$.

The terms in the forward scattering potential that depend on the magnetic field result from forward scattering off polarized background electrons and baryons that tend to have their spins anti-parallel to the magnetic field.  For $\nu_e$,
\begin{equation}
V_1^e (\rho, B, T; Y_e) = \left( - c_e + c_p + c_n \right) B,
\end{equation}
and for $\nu_\mu$ and $\nu_\tau$,
\begin{equation}
V_1 (\rho, B, T; Y_e) = \left( c_e + c_p + c_n \right) B .
\end{equation}
The difference between the forward scattering potentials for $\nu_e$ and $\nu_{\mu, \tau}$ is that $\nu_e$ have both charged and neutral current contributions from scattering off electrons, while $\nu_{\mu, \tau}$ have only neutral current contributions.  The energy scales in the protoneutron star are well below the threshold to thermally populate muon or tau states.

For degenerate electrons, $n_e \gg n_{e^+}$, so $n_e \approx Y_e n_b$.  The neutral current contribution from neutrino scattering off polarized electrons is
\begin{equation}
c_e B \approx 2.0 \,\rho_{14}^{1/3} B_{16} \left( \frac{Y_e}{0.04} \right)^{1/3} ~{\rm meV} ,
\end{equation}
for scattering off polarized protons is
\begin{equation}
c_p B \approx 0.85 \,\rho_{14} B_{16} \left( \frac{Y_e}{0.04} \right)  ~{\rm meV} ,
\end{equation}
and for scattering off polarized neutrons is
\begin{equation}
c_n B \approx 15 \, \rho_{14} B_{16} ~{\rm meV} ,
\end{equation}
where $B_{16}$ is the strength of the magnetic field in units of $10^{16}~{\rm G}$ \cite{als97}.

%The forward scattering potential for $\nu_\mu$ and $\nu_\tau$ is a function of the direction of its momentum vector relative to the magnetic field, while the forward scattering potential for $\nu_e$ is independent of the direction of the neutrino momentum.  The basis of this term comes from the fact that the magnetic field polarizes the background electrons to have their spins anti-parallel to the magnetic field.  As a result, the charged current and neutral contributions to the forward scattering potential of $\nu_e$ cancel out.  However, since the energy scales in the protoneutron star core are well below the threshold to thermally populate muon or tau states, there is a negligible charged current contribution to the forward scattering potential, resulting in a non-zero contribution to the $\nu_\mu$ and $\nu_\tau$ forward scattering potentials from this effect \cite{ec96}.

%Notice that the $\mu$ and $\tau$ neutrinos have a direction-dependent forward scattering potential, while the electron neutrinos do not.  This is the result of the polarization of electrons in the magnetic field, not from any neutrino magnetic moment, which we will neglect here.  

In this paper we will work with 2x2 mixing between $\nu_\mu$ and $\nu_s$ (and their $CP$-related counterparts).  This analysis will present a flavor of the physical mechanism that drives the pulsar kick through active-sterile neutrino transformation.  The effects of mixing with all three active species will be discussed in Section \ref{sec:caveats}.

%In order to create an anisotropic distribution of sterile neutrinos emitted from any given mass element, we should be concerned with $\mu$ or $\tau$ neutrinos since the forward scattering potential of these neutrinos, and as a result their resonant energy, depends on the direction between the local magnetic field and the direction of the neutrino momentum, $\cos \theta_\nu \propto \hat{\mathbf{k}} \cdot \vec{\mathbf{B}}$.  For concreteness, in the 2x2 mixing scheme discussed in this section, we will consider only mixing between $\nu_\mu$ and $\nu_s$ (and their $CP$-related counterparts).

At a given time and location in a neutron star, the local physical conditions (density, temperature, and magnetic field), along with the local chemistry (the $Y_\ell$), determine the resonant neutrino energy as a function of the neutrino direction relative to the magnetic field ($\cos \theta_\nu$).  The resonance condition, Eq.\ (\ref{eq:msw}), applies to $\nu_\mu \rightleftharpoons \nu_s$ when $V_\mu > 0$ or $\bar\nu_\mu \rightleftharpoons \bar\nu_s$ when $V_\mu < 0$.  

Sterile neutrinos are created when $\mu$ neutrinos coherently evolve through a resonance.  The resonant momentum is
\begin{equation}
p_{\rm res} = \frac{m_s^2}{2 V_\mu (\mathbf{r}, t; \cos \theta_\nu)} ,
\label{eq:resonance}
\end{equation}
where we have made the assumption that the vacuum mixing angle is small, and the tacit understanding that $p_{\rm res} > 0$ applies to $\nu_\mu \rightleftharpoons \nu_s$ and $p_{\rm res} < 0$ applies to $\bar\nu_\mu \rightleftharpoons \bar\nu_s$.  It is evident that the resonant momentum depends on the local conditions and the direction of the neutrino.

If the neutrino oscillation length at resonance is small compared to the width of the resonance, then the coherent evolution of the neutrino states is adiabatic resulting in the efficient conversion of active to sterile neutrinos (or anti-neutrinos).  That is, all neutrinos traveling in a given direction with the resonant momentum will be converted;  active neutrinos will become sterile, and any sterile neutrinos will be converted back to an active neutrino.  However, if the neutrino oscillation is of order the resonance width, or even larger, the neutrino propagation through the resonance is non-adiabatic and as a result the active-sterile neutrino conversion is less efficient, albeit non-zero.

$\nu_\mu$ or $\bar\nu_\mu$ with momenta dictated by the resonance condition are converted into sterile neutrinos with a probability dictated by how adiabatic the conditions are as the neutrino states cross the resonance.  The width of the resonance in momentum space is $p_{\rm res} \tan 2 \theta$, so a ``hole'' is created in the active neutrino and anti-neutrino distribution functions at the resonant momentum with the given width.  \ynm changes as neutrinos and anti-neutrinos are converted to sterile neutrinos and energy is radiated away as the sterile neutrinos freely stream out of the protoneutron star.

In the core of the protoneutron star, the degenerate neutrons are the dominant component in determining the local energy and pressure.  The leptons play a smaller role in the local energy and pressure budget of which the $\mu$-neutrinos and anti-neutrinos have only a fractional influence (less so early on when the electrons and electron-neutrinos are degenerate while the $\mu$-neutrinos are not).  So, it would be reasonable to proceed as if the production of sterile neutrinos have no feedback on the physical conditions of the neutron star.

Since the neutrons far outnumber the neutrinos, which in turn, far out number the resonant neutrinos that transform into sterile neutrinos, the concomitant change in \ynm is small.  The result is that the resonant momentum does not change significantly and if the hole in the neutrino distribution function is not repopulated, the sterile neutrino production mechanism would stall.  However, at the high densities found inside the protoneutron star core, scattering and neutron bremsstrahlung are efficient mechanisms for repopulating the hole created in the neutrino distribution function.  

As neutrinos are created at the resonant momentum, they are converted to sterile neutrinos as described above, resulting in continued change in \ynm.  Eventually, the change in \ynm will become large enough to shift the resonance to thermally populated regions of phase space and the process begins anew.  The active-sterile neutrino transformation resonance sweeps through the active neutrino distributions converting active neutrinos into sterile neutrinos.  The sweeping of the resonance continues until the forward scattering potential is altered so that the total active-sterile conversion converts an equal number of neutrinos as anti-neutrinos and thus \ynm is unchanged.

\section{Calculations}
\label{sec:calculations}

\subsection{Resonant Production}

As described in the previous section, the chemical evolution of a mass element in the protoneutron star is dominated by the scattering of neutrinos into the resonance.  When $\nu_\mu$ are scattered into the resonance and converted into $\nu_s$, \ynm decreases, and when $\bar\nu_\mu$ are scattered into the resonance and converted into $\bar\nu_s$, \ynm increases.  The evolution of \ynm is important in the determination of the spectra of sterile neutrinos emitted as a result of the active-sterile neutrino transformation.

The sterile neutrino production rate is equal to the rate active neutrinos are scattered into the hole in the neutrino distribution created by resonant active-sterile neutrino transformation.  The scattering rate of neutrinos into the hole in the distribution is proportional to the overall scattering rate.  In addition, the scattering rate is proportional to the size of the resonance, {\it i.e.},
\begin{equation}
\frac{d^3 p}{(2 \pi)^3} = \frac{p_{\rm res}^3 \tan 2 \theta \,d \cos \theta_\nu}{4 \pi^2} ,
\end{equation}
where, by symmetry, the azimuthal direction about the magnetic field is integrated out.  Finally, the scattering rate is proportional to the fraction of neutrinos available to scatter into the resonant momentum.  Thus, the sterile neutrino production rate for neutrino states with momenta in the direction described by $\cos \theta_\nu$ is
\begin{equation}
\frac{d n_{\nu_s}}{d t} d  \cos \theta_\nu \approx \frac{p_{\rm res}^3 \tan 2 \theta \,d \cos \theta_\nu}{4 \pi^2} \frac{p_{\rm res}^2 / T^2}{e^{p_{\rm res}/T} + 1} \Gamma_{\nu_\mu} \mathcal{P}_{\nu_\mu \rightarrow \nu_s},
\end{equation}
where $\Gamma_{\nu_\mu}$ is the total $\nu_\mu$ scattering rate and $\mathcal{P}_{\nu_\mu \rightarrow \nu_s}$ is the probability that a $\nu_\mu$ with the resonant momentum will transform to $\nu_s$ as it traverses the resonance.

In the core of the protoneutron star, the neutrino scattering rate is primarily due to neutrino-nucleon scattering \cite{iank89}, so the total neutrino scattering rate is
\begin{equation}
\Gamma_{\nu_\mu} \sim \frac{\rho}{m_n} G_F^2 p^2 .
\end{equation}
At the temperatures of interest, muons are not thermally populated, so neutral current interactions are responsible for the scattering of the $\nu_\mu$ and $\bar\nu_\mu$ into the resonant momentum states.  Consequently, the production rates for $\bar\nu_\mu \rightarrow \bar\nu_s$  should have the same form as those for $\nu_\mu \rightarrow \nu_s$ shown above.

The probability of conversion should be unity when the neutrino oscillation length is small compared to the resonance width ($\ell_{\rm res} \ll \delta t$) and tends toward zero when $\ell_{\rm res} \gg \delta t$.  Here, we use the Landau-Zener jump probability \cite{landau, zener}, 
\begin{equation}
\mathcal{P}_{\nu_\mu \rightarrow \nu_s} \approx \mathcal{P}_{\bar\nu_\mu \rightarrow \bar\nu_s} \approx 1 - e^{-\pi \gamma / 2} ,
\end{equation}
which approaches unity for $\gamma \gg 1$ and is proportional to $\gamma$ for $\gamma \ll 1$.  The adiabaticity parameter is
\begin{equation}
\gamma \equiv \frac{\delta t}{\ell_{\rm res}} = \frac{m_s^2}{2 p} \left\vert \frac{1}{V_\mu} \frac{d V_\mu}{d x} \right\vert^{-1} \sin 2 \theta .
\end{equation}

The evolution of \ynm is thus dictated by
\begin{equation}
\frac{d \ynm}{d t} = - \frac{m_n}{\rho} \int_{-1}^{+1} \left( \frac{d n_{\nu_s}}{d t} - \frac{d n_{\bar\nu_s}}{d t} \right) d \cos \theta_\nu .
\label{eq:dydt}
\end{equation}
The evolution of \ynm is important because the value of \ynm along with the other lepton numbers and the physical conditions in the protoneutron star determine the resonant momenta.

The sterile neutrinos that are created in these resonances subsequently stream freely out of the protoneutron star since they no longer interact with the dense environment.  By symmetry, the total momentum of these sterile neutrinos are either parallel or anti-parallel to the magnetic field.  Thus, in the calculation of the total momentum emitted in sterile neutrinos, we need only take the component of the neutrino momentum along the magnetic field into account when calculating the total momentum emitted from the protoneutron star.  The rate at which momentum is emitted in sterile neutrinos per unit mass is 
\begin{equation}
\frac{d p_s}{d m \, dt} = \frac{1}{\rho}  \int_{-1}^{+1} \left( \frac{d n_{\nu_s}}{d t} + \frac{d n_{\bar\nu_s}}{d t} \right) p \cos \theta_\nu \,d \cos \theta_\nu .
\label{eq:dpdt}
\end{equation}
The rate at which energy is emitted in sterile neutrinos per unit mass is
\begin{equation}
\frac{d E_s}{d m \, dt} = \frac{1}{\rho}  \int_{-1}^{+1} \left( \frac{d n_{\nu_s}}{d t} + \frac{d n_{\bar\nu_s}}{d t} \right) p \,d \cos \theta_\nu .
\label{eq:dEdt}
\end{equation}

The calculation presented will consider the sterile neutrino production at a location in a neutron star with a fixed value of its density, $\rho$ and temperature $T$.  The initial conditions on the lepton numbers are chosen to be $Y_e = 0.04$, $Y_{\nu_e} = 0.07$, and $Y_{\nu_\mu} = Y_{\nu_\tau} = 0$. 

%The electrons are relativistic and degenerate, so $n_e \gg n_{e^+}$, and $n_e \approx Y_e n_b$.  Thus, the forward scattering potential is
%\begin{equation}
%V_\mu (Y_{\nu_\mu}, \cos \theta_\nu) = - V_0 (\rho) f_\mu (Y_\ell) + V_1 (\rho, B; Y_e) \cos \theta_\nu ,
%\end{equation}
%with $V_0 (\rho) \approx 3.8 \rho_{14}~{\rm eV}$, and
%\begin{equation}
%V_1 (\rho, B; Y_e) \approx 4.0 \times 10^{-3} \rho_{14}^{1/3} B_{16}~{\rm eV} \left( \frac{Y_e}{0.04} \right)^{1/3} ,
%\end{equation}
%where $\rho_{14}$ is the density in units of $10^{14}~{\rm g}/{\rm cm}^3$, and $B_{16}$ is the strength of the magnetic field in units of $10^{16}~{\rm G}$.

%With the initial conditions listed above, $f_\mu$ is positive, which means that $V_\mu$ is negative for a range of the allowed values of $\cos \theta_\nu$.  For this range of directions, $\bar\nu_\mu \rightarrow \bar\nu_s$ will occur.  If the magnetic field is sufficiently large compared to the density ({\it i.e.}, $B_{16} \gtrsim 5.5 \rho_{14}^{2/3}$), then $\nu_\mu \rightarrow \nu_s$ will occur for neutrinos directed in the same direction as the magnetic field.  The form of the resonance condition, Eq.\ (\ref{eq:resonance}), tends to have higher resonance energies for neutrinos directed into the hemisphere that is parallel to the magnetic field (hereafter, the ``northern hemisphere'') than the hemisphere anti-parallel to the magnetic field (hereafter, the ``southern hemisphere''), resulting in an anisotropy of the emitted momentum in sterile neutrinos.  

With the initial conditions listed above, $f_\mu$ is positive, which means that $V_\mu$ is negative for all allowed values of $\cos \theta_\nu$ as long as $B < 8 \rho_{14}^{2/3} \times 10^{18}~{\rm G}$.  The form of the resonance condition, Eq.\ (\ref{eq:resonance}), tends to have higher resonant momenta for neutrinos directed into the hemisphere parallel to the magnetic field (hereafter, the ``northern hemisphere'') than the hemisphere anti-parallel to the magnetic field (hereafter, the ``southern hemisphere''), resulting in an anisotropy of the emitted momentum in sterile neutrinos.

\subsection{Reabsorption}

The same MSW-like resonance that transforms active neutrinos into sterile neutrinos will also transform sterile neutrinos back into active neutrinos if the neutrino's momentum is once again resonant at a later point in time along its worldline.  As a consequence, the total momentum in sterile neutrinos that escape from the neutron star is reduced as sterile neutrinos that would otherwise escape the star are converted back into active neutrinos and are rethermalized in the sea of active neutrinos in another location.  The result is that evolution of $Y_{\nu_\mu}$ is altered in a manner that depends on the location of the mass element in the star and on the regions of phase space in the $\nu_\mu$ and/or $\bar\nu_\mu$ distribution functions that are reabsorbed.

%The magnitude of the effect of reabsorption of sterile neutrinos is not a simple calculation.  
%%Even if we were to assume a spherically symmetric neutron star, $Y_{\nu_\mu}$ would not be similarly spherically symmetric.  
%The evolution of $Y_{\nu_\mu}$ depends not only on the number density of active neutrinos that are converted into sterile neutrinos, but also on the number of incoming sterile neutrinos with the resonant momentum (both magnitude and direction).  As a result, the evolution of $Y_{\nu_\mu}$ is different for different regions in the neutron star, even if the density and temperature is the same in those regions.  

In a mass element of the star, the sterile neutrinos that are emitted in the hemisphere directed out of the star will not traverse another resonance and will thus escape from the star.  On the other hand, the sterile neutrinos that are emitted in the hemisphere directed toward the center of the star will encounter another resonance and thus be reabsorbed in another mass element on the other side of the star.  The many-decades range of densities (and relatively small range of $f_\mu$) in the core of the protoneutron star effectively assures that inward emitted neutrinos will encounter another resonance.  (More general geometries may result in multiple resonances along the trajectory of a sterile neutrino.  In such a case, the interior resonances will have a reabsorption rate comparable to the emission rate because there will be resonant neutrinos in both inward and outward directions.  Thus, only the outermost resonances will contribute to the overall effect.)

Roughly, the global kinematic effect of this description is that only the sterile neutrinos emitted out of the star escape from the star because these neutrinos are not reabsorbed.  On the other hand, the local effects on energy transport and the evolution of \ynm depend only on the sterile neutrinos emitted into the star, because a similar number of neutrinos are emitted in an outward direction as are reabsorbed as neutrinos traveling in the same direction, but have traversed the center of the star.
%Locally, the reabsorption of sterile neutrinos results in a reduction of the cooling rate of the mass element.  Globally, the reabsorption of sterile neutrinos weakens the strength of the pulsar kick by active-sterile neutrino transformation.  
To properly determine the transport of energy from one location in the protoneutron star to another through the transformation, $\nu_\mu \rightarrow \nu_s \rightarrow \nu_\mu$, is a complex, multi-dimensional ray tracing transport problem.  (Refs.\ \cite{hf06, hf07} considered a similar energy transport scheme in one dimension.)  

In a mass element of the protoneutron star, the number of sterile neutrinos that are reabsorbed is roughly half the number of sterile neutrinos emitted.  Thus, it is a good approximation to halve the rate of change of $Y_{\nu_\mu}$ in Eq.\ (\ref{eq:dydt}). The momentum and energy rates emitted in sterile neutrinos can be estimated by averaging the output of diametrically opposed mass elements.  Consequently, the momentum and energy emission rates in Eqs.\ (\ref{eq:dpdt}) and (\ref{eq:dEdt}) are also roughly halved.  The results in the following section will be presented with respect to these reabsorption effects on the chemical and kinematic evolution and will employ the approximations stated above.

%However, estimates can be made to the magnitude of this effect.  Only outward directed neutrinos escape the star, so roughly one-half of the emitted sterile neutrinos will be absorbed elsewhere in the protoneutron star.  Thus, both the energy and momentum emission from the protoneutron star in the form of sterile neutrinos will be roughly one-half as large as predicted in the previous section, solely from the reabsorption of sterile neutrinos.  Similarly, the local energy transport of sterile neutrinos from a mass element is roughly one half of that predicted earlier in this paper.
%However, the effects on the evolution of $Y_{\nu_\mu}$ must be taken into consideration because the rate of momentum emitted in sterile neutrinos depends on the local value of $Y_{\nu_\mu}$.  

\section{Results}
\label{sec:results}

\subsection{An Example}

\begin{figure*}
  \includegraphics[width = 4in, angle = 270]{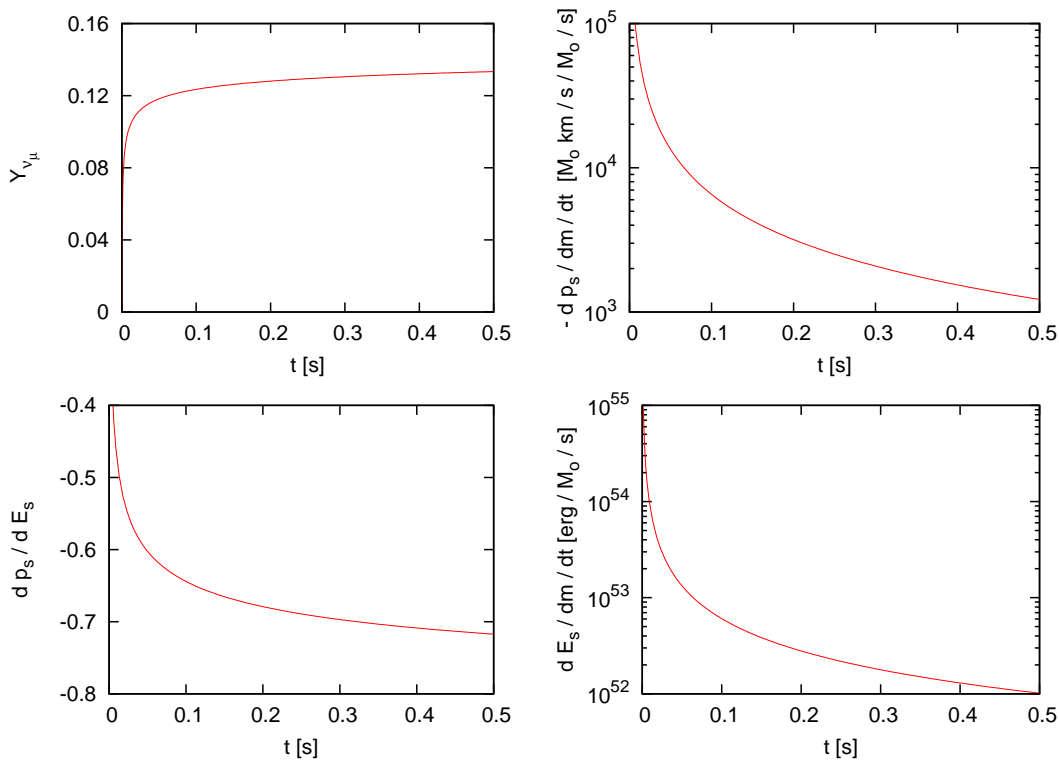}
  \caption{\label{fig:0.02-10-20} Results for the calculation of the production of sterile neutrinos in a mass element of the protoneutron star.  Upper left panel is the temporal evolution of $Y_{\nu_\mu}$.  Upper right panel is the rate per unit mass of momentum of sterile neutrino produced in the opposite direction of the magnetic field.  Lower left panel is the ratio of the rates of net momentum along the magnetic field to energy emitted in sterile neutrinos.  Lower right panel is the rate per unit mass of energy emitted in sterile neutrinos.  Sterile neutrino parameters:  $m_s = 5~{\rm keV}$, $\sin^2 2 \theta = 10^{-9}$.  Physical parameters:  $B_{16} = 10$, $T = 20~{\rm MeV}$, $\rho_{14} = 0.02$, $\vert d \log V_\mu / d r \vert^{-1} = 10~{\rm km}$.  Initial conditions:  $Y_e = 0.04$, $Y_{\nu_e} = 0.07$, $Y_{\nu_\mu} = Y_{\nu_\tau} = 0$.}
\end{figure*}

As an example of the the sterile neutrino production mechanism discussed earlier, we focus on a mass element in a neutron star with $T = 20~{\rm MeV}$, and $B = 10^{17}~{\rm G}$.  The initial conditions stated in the previous section on the lepton numbers are also employed:  $Y_e = 0.04$, $Y_{\nu_e} = 0.07$, and $Y_{\nu_\mu} = Y_{\nu_\tau} = 0$.  Finally, the density scale height, $\vert d \log V_\mu / d r \vert^{-1} = 10~{\rm km}$, is used in calculating the adiabaticity parameter.  

Throughout this section, we consider the parameters of the mass element in the protoneutron star remain unchanged throughout the calculation.  While this may be a good assumption on the collisional timescale, as mentioned in the previous section, on the longer dynamical timescales considered in the calculation, these assumptions ought to break down.  A discussion is presented in Section \ref{sec:caveats} to qualitatively discuss consequences of a multitude of effects that may be present if this mechanism were to be included in a self-consistent simulation of the formation of a pulsar.

Figure \ref{fig:0.02-10-20} summarizes the chemical and kinematic evolution due to the production of sterile neutrinos of a mass element in the protoneutron star with these physical characteristics and a density of $2 \times 10^{12}~{\rm g}/{\rm cm}^3$.  The upper left graphs shows the evolution of \ynm.  The two graphs on the right show the rate per unit mass of the momentum of the emitted sterile neutrinos  (upper) and energy of sterile neutrinos emitted (lower).  The lower left graph shows the ratio of these momentum and energy rates, a measure of the asymmetry in the sterile neutrino emission.  If this ratio [$d p_s / d E_s = (d p_s / dm / dt) / (d E_s / dm / dt)$] is equal to -1, then the sterile neutrinos are emitted solely opposite to the direction of the magnetic field.

%\begin{figure}
% (a) \includegraphics[width = 2in, angle = 270]{ynu-0.12-3-15.ps}
% (b)\includegraphics[width = 2in, angle = 270]{rate-0.12-3-15.ps}
%\caption{\label{fig:0.12-3-15} (a)  The temporal evolution of $Y_{\nu_\mu}$.  (b) The rate per unit mass of momentum of sterile neutrinos produced along the direction of the magnetic field (solid curve) and the ratio of the rates of net momentum along the magnetic field to energy emitted in sterile neutrinos (dashed curve).  Sterile neutrino parameters:  $m_s = 5~{\rm keV}$, $\sin^2 2 \theta = 10^{-9}$.  Physical parameters:  $B_{16} = 3$, $T = 15~{\rm MeV}$, $\rho_{14} = 0.12$, $\vert d \log V_\mu / d r \vert^{-1} = 10~{\rm km}$.  Initial conditions:  $Y_e = 0.04$, $Y_{\nu_e} = 0.07$, $Y_{\nu_\mu} = Y_{\nu_\tau} = 0$.}
%\end{figure}

%One thing to notice from Figure \ref{fig:0.12-3-15}b is that $d p_s / dm / dt$ is nearly proportional to $d p_s / d E_s$.  One can determine the value of the energy rate of sterile neutrino emission from 
%\begin{equation}
%\frac{d E_s}{dm \,dt} = \frac{d p_s}{dm \,dt} \times \left( \frac{d p_s}{d E_s} \right)^{-1} .
%\end{equation}
%Hence, the energy rate of sterile neutrino emission is roughly constant over the time period shown and is approximately equal to $1.6 \times 10^{54}~{\rm erg}/{\rm s}/M_\odot$.

The general shapes of the curves in Figure \ref{fig:0.02-10-20} are consequences of the mathematical form of the forward scattering potential and the resonances in this problem.  There are three generic features of the sterile neutrino production mechanism discussed in this paper.  (1) \ynm increases from its initial value of 0 and asymptotes to a fixed value that depends on the initial value of the lepton numbers in electrons and electron neutrinos.  (2) The net momentum of the emitted sterile neutrinos is opposite the direction of the magnetic field.  (3)  The magnitude of the momentum and energy emission rates is initially large, but is suppressed by multiple orders of magnitude on a timescale of a fraction of a second.  
%(4)  The asymmetry of the sterile neutrino emission grows as the magnitude of the emissions decrease.

%
%(2) The rate at which momentum emitted in sterile neutrinos typically starts at a large positive value and asymptotes to zero as \ynm asymptotes to its final value.  (3)  The energy rate of sterile neutrino emission is roughly constant.

\begin{figure}
 \includegraphics[width = 2in, angle = 270]{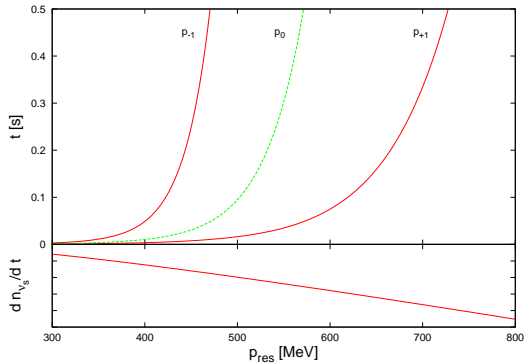}
\caption{\label{fig:fnu-0.02-10-20} The evolution of resonant momenta $p_{+1}$, $p_{-1}$ and $p_0$ (upper graph) using the same parameters as in Figure \ref{fig:0.02-10-20}.  $p_{+1}$ ($p_{-1}$) is the resonant momentum for neutrinos traveling parallel (anti-parallel) to the magnetic field.  $p_0$ is the resonant momentum for neutrinos traveling perpendicular to the magnetic field.  The lower graph is the sterile neutrino production rate (each tic represents two decades difference in the production rate).}
\end{figure}

These characteristics of the active-sterile transformation mechanism in the protoneutron star core can be explained by understanding the regions in phase space where the transformation is taking place.  Figure \ref{fig:fnu-0.02-10-20} shows the evolution of three important resonant momenta through the first half-second of the process.  $p_{+1}$ ($p_{-1}$) represents the resonant momentum for neutrinos that are traveling parallel (anti-parallel) to the magnetic field.  (Thus, $\cos \theta_\nu = +1(-1)$.)  In a similar vein, $p_0$ represents the resonant momentum of neutrinos that are traveling perpendicular to the magnetic field.  The lower figure is the sterile neutrino production rate in a logarithmic scale with each tic mark representing a difference in the rate of two orders of magnitude.   The basis of the emission asymmetry can be seen as the sterile neutrino rate for neutrinos emitted into the southern hemisphere is orders of magnitude larger than the rate for those emitted into the northern hemisphere.

In Figure \ref{fig:fnu-0.02-10-20}, $\bar\nu_\mu$ are converted to $\bar\nu_s$ into the southern hemisphere with momenta between $p_{-1}$ and $p_0$.  On the other hand, $\bar\nu_\mu$ are converted to $\bar\nu_s$ into the northern hemisphere with momenta between $p_0$ and $p_{+1}$.  

\ynm increases because $\bar\nu_\mu$ are transformed into sterile neutrinos, while the $\nu_\mu$ are not affected by this resonant conversion.  As \ynm increases, $f_\mu (Y_\ell)$ decreases, resulting in an increase in each of the respective resonant momenta.  As the resonant momenta shift higher, the sterile neutrino production rates become suppressed both exponentially due to the paucity of active neutrinos with these high energies available to scatter into the resonance and linearly as the conversion becomes non-adiabatic.

The asymmetry in the overall momentum of the emitted sterile neutrinos is because $p_{-1} < p_{+1}$.  As is demonstrated in Figure \ref{fig:fnu-0.02-10-20}, the lower resonant momenta correspond to much larger sterile neutrino emission rates.  Thus, there are more sterile neutrinos emitted into the southern hemisphere than the northern hemisphere, resulting in a net momentum pointing anti-parallel to the magnetic field. As $\bar\nu_\mu$ are converted into sterile neutrinos, causing the increase in \ynm as discussed previously, the resonant momenta become larger where the production rate decreases exponentially resulting in the steep decline in the rates of momentum and energy emitted in sterile neutrinos as seen in Figure \ref{fig:0.02-10-20}.

\subsection{Generating the pulsar kick and cooling the star}

\begin{figure}
 \includegraphics[width = 2in, angle = 270]{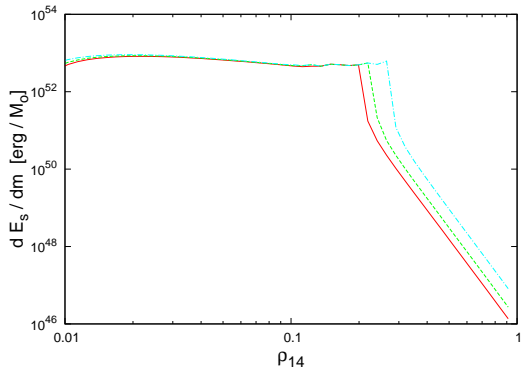}
\caption{\label{fig:rho-10-20} The energy per unit mass of sterile neutrino emission as a function of density integrated over the first $0.5~{\rm s}$ (solid), $1~{\rm s}$ (dashed), and $t = 3~{\rm s}$ (dot-dashed).  All the other physical and neutrino properties and initial conditions are the same as in Figure \ref{fig:0.02-10-20}.  }
\end{figure}

Figure \ref{fig:rho-10-20} illustrates the energy emission per unit mass of sterile neutrinos as a function of density assuming $T = 20~{\rm MeV}$ and $B_{16} = 10$.  The curves are the total energy emitted in the first half-second, the first second, and the first three seconds.  It can be seen that most of the sterile neutrino emission occurs early on in the process, as longer integration times result in disproportionately smaller increases in the overall energy emission.

%To determine the effect of the mechanism described above, the effective mass of the region in the protoneutron star where this effect occurs must first be estimated.  Figure \ref{fig:rho-3-15} illustrates the energy rate of sterile neutrino emission as a function of density assuming the same parameters discussed above, to wit, $T = 15~{\rm MeV}$ and $B_{16} = 3$.  The three curves are for the energy rate at $0~{\rm s}$, $1~{\rm s}$ and $3~{\rm s}$.

%Initially, for $\rho_{14} \lesssim 0.4$, the energy emission rate of sterile neutrinos is roughly constant.  This is the regime expected by requiring initially that both $\bar\nu_\mu$ and $\nu_\mu$ to be transformed into sterile states.  Higher densities suffer from the smaller resonant phase space for neutrino transformation discussed in the previous section when $\bar\nu_\mu$ have resonant transitions while $\nu_\mu$ do not.

%At later times, the transition between the two regimes moves to higher densities.   If $\ynm > 0$, then the condition for both $\bar\nu_\mu$ and $\nu_\mu$ to be converted to sterile states is
%\begin{equation}
%B_{16} \gtrsim 6.7 f_\mu (Y_\ell) \rho_{14}^{2/3} .
%\end{equation}
%In these higher densities, $\bar\nu_\mu$ are initially transformed but $\nu_\mu$ are not, resulting in an increase in \ynm.  Increasing \ynm leads to a decrease in $f_\mu (Y_\ell)$.  The result is that as \ynm increases, higher densities will have the large energy and momentum emission rates that are characteristic of having both \bnm and \nm transforming to sterile states.

\begin{figure}
 \includegraphics[width = 2in, angle = 270]{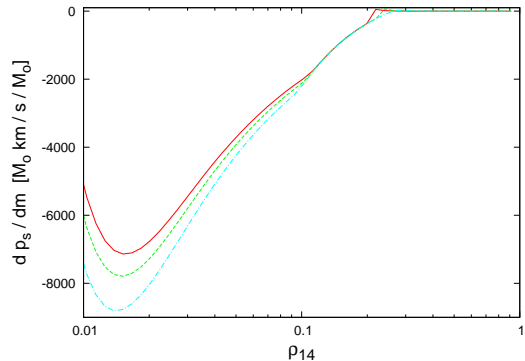}
\caption{\label{fig:mom-10-20} The total momentum per unit mass of sterile neutrinos emitted from a mass element in the protoneutron star as a function of density at times $t = 0.5~{\rm s}$ (solid), $t = 1~{\rm s}$ (dashed), and $t = 3~{\rm s}$ (dot-dashed). All the other physical and neutrino properties and initial conditions are the same as in Figure \ref{fig:0.02-10-20}.  }
\end{figure}

Figure \ref{fig:mom-10-20} shows the output in terms of the net momentum in sterile neutrinos emitted from the protoneutron star.  The curves show the total momentum per unit mass that is emitted in the form of sterile neutrinos from the protoneutron star as a function of the local density at three different times in the evolution.  Once again, it can be seen that most of the sterile neutrino emission occurs early on in the process.  The momentum output is negative because the net momentum emitted is anti-parallel to the magnetic field as discussed above.

%For densities $0.04 \lesssim \rho_{14} \lesssim 0.7$, a large momentum is emitted in sterile neutrinos is generated in the first few seconds.

%For the sterile neutrino parameters and physical conditions calculated in the example above, sterile neutrinos remove energy from the protoneutron star at a rate of $\sim 10^{54}~{\rm erg}/M_\odot/{\rm s}$ for $\rho_{14} \lesssim 0.6$.  If we estimate the mass of the region in the protoneutron star with $\rho_{14} \lesssim 0.6$ from supernova simulations ({\it e.g.}, Ref.\ \cite{sr05}), the total mass involved in the mechanism is on the order of $\mbox{a few} \times 10^{-2}~M_\odot$.  Thus, if the emission is integrated over the entire lifetime of the collapse, roughly $10~{\rm s}$, the total energy emitted in sterile neutrinos is $\sim 10^{53-54}~{\rm erg}$.

To determine the overall magnitude of the effect, the results from Figures \ref{fig:rho-10-20} and \ref{fig:mom-10-20} must be integrated over a density profile for the protoneutron star.  Estimating the density profile from Ref.\ \cite{sr05}, the total energy emitted is $\sim 6 \times 10^{52}~{\rm erg}$ and the resultant pulsar kick velocity is $\sim 2000~{\rm km}/{\rm s}$, opposite the direction of the magnetic field.  

While the strength of the pulsar kick is larger than necessary to explain the observed pulsar velocities, there are other effects -- discussed below -- that may decrease the magnitude of the pulsar kick.
The observation of active neutrinos from SN1987A implies that the total energy emitted in sterile neutrinos should be $\lesssim  10^{53}~{\rm erg}$ \cite{ss87, gaan88}, lest the emission of sterile neutrinos overcool the protoneutron star, reducing the energy budget left for active neutrino emission to match observations.  It appears that the active-sterile neutrino transformation mechanism presented can generate a sufficient kick to explain the observational data, yet not overcool the protoneutron star.

\subsection{Other Considerations}
\label{sec:caveats}

The calculation performed above involved the determination of the net momentum emitted in sterile neutrinos from a volume element in a protoneutron star.  Although the reabsorption of sterile neutrinos due to subsequent traversal of resonances is a complex process, the calculation presented above attempted to account for this effect by self-consistently reducing the sterile neutrino emission rates.
%In the calculation presented, the estimated energy emitted is larger than current constraints by less than an order of magnitude, and the estimated momentum emitted is about an order of magnitude larger than the momenta of the fastest observed pulsars.  
However, there are three other considerations to take into account when determining the magnitude of the active-sterile neutrino transformation mechanism for large pulsar kicks:  dynamical evolution of the physical properties of the mass element, the dynamical evolution of the active neutrino seas, and sterile neutrino mixing with all three active flavors.

To accurately model the momentum generated in sterile neutrinos, this mechanism for generating sterile neutrinos through active-sterile neutrino transformation would need to be calculated alongside a supernova simulation.  The estimates presented in this paper suggest that this mechanism can provide a sufficient kick yet not overcool the star.  

%In addition to the dynamical effects of the collapsing protoneutron star, 
The interplay between the emission of a significant fraction of the binding energy of the star and the dynamical evolution of the collapsing star can, in turn, affect the total sterile neutrino output.  The temperature of the mass element has a strong leverage on the output of sterile neutrinos.
%The most significant factor that will affect the output in sterile neutrinos is the evolution of the temperature of the mass element.  
%Additionally, if the sterile neutrinos begin to remove a significant fraction of the energy budget of the mass element, the temperature should decrease.  
By inspection of the sterile neutrino production rate, the total energy output rate in sterile neutrinos is proportional to $T^{5-6}$ (5 if most of the sterile neutrinos are produced non-adiabatically and 6 if they are produced adiabatically).  Thus, small changes in temperature can result in significant changes in the energy and momentum emission rates.  However, one must be cognizant not to overcool the star with respect to the inferred temperature of the neutrinosphere from SN1987A \cite{ss87, gaan88}.  

Even without the emission of sterile neutrinos, the temperature of a mass element in the protoneutron star will evolve.  At the densities of interest for the calculation presented above (just within the neutrinosphere), the temperature will decrease on a roughly 10 second timescale as the heat generated by the infall and bounce stages of the supernova are both emitted through the neutrinosphere and diffuse toward the center of the protoneutron star \cite{prplm99}.  
%In addition, the emission of sterile neutrinos should also decrease the temperature, because if the temperature remained constant, the total emission in sterile neutrinos would rival the total energy budget of the star.
The emission of sterile neutrinos will affect the temperature profile of the protoneutron star at late times as the total energy emitted becomes a more significant fraction of the total energy budget.  In addition, the mechanism that generates the large magnetic fields necessary for this effect could decay as energy is removed by sterile neutrino emission.

Most of the sterile neutrino emission occurs in the first $\sim 50~{\rm ms}$.  In half a second, roughly $80\%$ of the total momentum is emitted (assuming unchanging temperature and magnetic field).  The favorable conditions of temperature and magnetic field presented in this work need only be present for a fraction of a second in order to produce a kick consistent with observations.  Hence, the evolution of the temperature and magnetic field strength is unlikely to affect the strength of the pulsar kick generated by this mechanism beyond the tens of percent level.

%The emission of sterile neutrinos will affect the temperature profile of the protoneutron star at late times ($\gtrsim {\rm a\, few}~{\rm s}$), when the total energy emitted becomes comparable to the total energy budget.  The strong dependence of the emission rates on the temperature mean that a small decrease in temperature can correspond to a larger reduction in the cooling rate due to sterile neutrinos.  Figure \ref{fig:0.12-3-15} suggests that this will have a much smaller effect on the total momentum output as compared to the total energy output because the momentum output decreases with time as $\ynm \rightarrow Y_{\nu_\mu,{\rm eq}}$, while the energy output is roughly constant.

%In addition, as the emission of sterile neutrinos cool the protoneutron star, the dynamo mechanism that generates the large magnetic field necessary in this mechanism may be affected.  The additional loss of energy due to sterile neutrino emission could reduce the strength of the dynamo mechanism, in turn resulting in reduced magnetic fields throughout the star.  As a result both the net momentum and energy emitted in sterile neutrinos decline with decreasing magnetic field strength.  As reasoned above, this will reduce the energy output significantly more than it will reduce the momentum output.

A number of dynamical effects can alter the results presented here for a static region.  As the protoneutron star collapses, the density scale height throughout the star will be reduced.  As a result, the adiabaticity parameter will proportionally shrink, reducing the number of sterile neutrinos created, especially at the highest energies.  The dynamical timescale is likely to be similar to or longer than the timescale for changes in the temperature and magnetic field, so these dynamic effects will reduce the strength of the kick, albeit less so than changes in the temperature and magnetic field.  Gradients in density and temperature will result in gradients in \ynm which will lead to diffusion that will act to reduce the disparity of \ynm in the protoneutron star.  However, the neutrino diffusion timescale is likely longer than the peak sterile neutrino emission timescale, so the effects of this diffusion are minimal in changing the mechanism.
%Finally, gradients in density and temperature will result in gradients in \ynm which will lead to diffusion that will act to reduce the disparity of \ynm in the protoneutron star.  The result is that it will take longer to attain the equilibrium value of \ynm, which affects the output in momentum -- typically increasing the net momentum output -- but having little effect on the energy output.

In this calculation, it was assumed that there was only vacuum mixing between the sterile neutrino and $\nu_\mu$.  However, if the sterile neutrino mixes with one active neutrino, it would likely also mix with the other active neutrinos as well.  Each active neutrino species will undergo active-sterile neutrino transformation in a similar fashion to the mechanism described for $\nu_\mu$.  Each will exhibit a similar asymmetry in the momentum of emitted neutrinos resulting from the term in the forward scattering potential that depends on both the strength and direction of the magnetic field.  All three flavors of active anti-neutrinos will be resonantly transformed into sterile neutrinos leading resulting in the increase in each respective lepton number. Since each flavor's lepton number will increase, the associated increase of the resonant momenta (as illustrated in Figure \ref{fig:fnu-0.02-10-20}) will be more rapid.  The result is the combinations of two differing effects:  strengthened by all three flavors being involved in the transformation, yet weakened because the sterile neutrino emission rates are suppressed as the resonant momenta increase.
%Active-sterile neutrino transformations involving $\nu_\tau$ will exhibit a similar asymmetry in the momentum of the emitted neutrinos, while the active-sterile transformations involving $\nu_e$ will not exhibit the asymmetry because the forward scattering potential for electron neutrinos is independent of the magnetic field.  Nevertheless, as the active-sterile neutrino transformation process occurs in all three active neutrino flavors, the corresponding $Y_\nu$ for each flavor increases.  The result is that it will take less time to attain an equilibrium value of \ynm.  At the equilibrium value, the momentum asymmetry vanishes, so this effect would reduce the magnitude of the pulsar kick while leaving the energy emission largely unaffected.

If $\nu_e$ mixes with the sterile neutrino, then a delayed kick, urca production mechanism can occur \cite{fkmp03, kmm08}.  In this urca production mechanism, the standard urca reactions, $e^- + p \rightleftharpoons n + \nu_e$ and $e^+ + n \rightleftharpoons p + \bar\nu_e$, can produce sterile neutrinos instead of electron neutrinos with a rate suppressed by $\sin^2 2 \theta_m$.  The neutrino states produced will have an asymmetry with more neutrinos created anti-parallel to the magnetic field than parallel to the field \cite{drt85}.  Thus, any contribution from the urca mechanism will augment a pulsar kick generated by  the active-sterile neutrino transformation mechanism.
%As a result, the pulsar kick generated by the urca mechanism opposes the kick created by the active-sterile neutrino transformation mechanism, resulting in a weaker kick.

In the urca mechanism, non-resonant active-sterile neutrino transformation in the core requires a few seconds for $V_e \rightarrow 0$.  Once the forward scattering potential for $\nu_e$ reaches zero, the non-resonant sterile neutrino production rate is maximized \cite{fkmp03, kmm08}.  The magnitude of this effect depends the conditions in throughout the core of the protoneutron star many seconds after the active-sterile neutrino transformation mechanism was initiated.  The large emission of energy in sterile neutrinos during the active-sterile neutrino transformation mechanism phase will affect the temperature and magnetic field strength throughout the core, reducing the available energy budget for further emission in sterile neutrinos.

\section{Discussion}
\label{sec:discussion}

The active-sterile neutrino transformation mechanism discussed in this paper can feasibly provide the requisite pulsar kick to explain the highest observed pulsar velocities.  It is interesting that the phase space of sterile neutrino parameters (mass and mixing angle with active neutrinos) where active-sterile neutrino transformation can explain pulsar kicks overlaps with the phase space of sterile neutrino parameters that provide a viable dark matter candidate (see Ref.\ \cite{kusenko09}).  It is enticing that two astrophysical phenomena could be explained by the same sterile neutrino, a straight-forward extension to the Standard Model.  Combined with tantalizing preliminary evidence for sterile neutrino dark matter that is consistent with this overlap in phase space \cite{lk10} makes for an interesting window in sterile neutrino physics.

Although several mechanisms propose to explain pulsar kicks without the introduction of sterile neutrinos, there are deficiencies that plague these mechanisms.  The evolution of close binaries \cite{ggo70} and the asymmetric emission of electromagnetic radiation \cite{ht75} have been proposed to solve the pulsar kick problem, but neither provide the requisite kick to explain the largest pulsar velocities.

Hydrodynamic mechanisms for generating pulsar kicks during the supernova collapse can result in pulsar velocities of a few hundred km/s, comparable to the average observed velocities, but it remains to be seen if they can explain the largest observed pulsar velocities \cite{nbblo10, spjkm04, skjm06, wjm10}.  In addition, it is unclear whether the observed correlation between pulsar velocity and spin axis \cite{johnston05, wlh06, nr07, nr08} can be explained with hydrodynamic mechanisms in three-dimensional simulations.  The active-sterile neutrino transformation mechanism discussed in this paper would be consistent with the velocity-spin correlation because although the generated pulsar kick is parallel to the magnetic field, any discrepancy between the magnetic field and protoneutron star spin will be averaged out due to the rotation of the star \cite{kusenko09}.

The considerations presented in this work do not change the general range in sterile neutrino parameter space (mass and mixing angle) from past work on this mechanism.  Requiring that the evolution of the neutrinos are both coherent and adiabatic constrain the sterile neutrino parameter space available to provide pulsar kicks through resonant active-sterile neutrino transformation.  This parameter space can be found in Refs.\ \cite{kusenko09, lk10, fkmp03, kmm08}.

\begin{figure}
 \includegraphics[width = 2in, angle = 270]{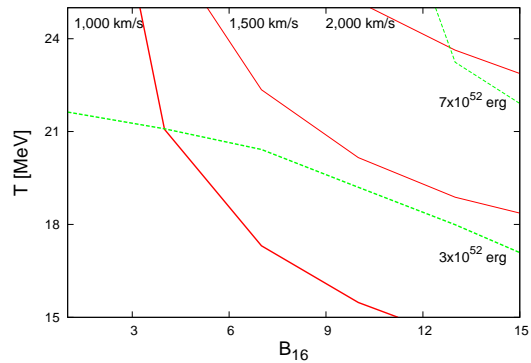}
\caption{\label{fig:BT} Estimates of the total energy emitted in sterile neutrinos and the magnitude of the pulsar kick created by active-sterile neutrino transformation mechanism in the parameter space of magnetic field ($B_{16}$) and temperature.  The protoneutron star parameters are those in the relevant densities ($10^{-2} \lesssim \rho_{14} \lesssim 1$) for the mechanism.  The solid lines are contours of the magnitude of the pulsar kick, while dashed lines are contours of the energy emission.}
\end{figure}

This work can put some basic constraints on the physical conditions within the core of the protoneutron star where the active-sterile neutrino transformation is taking place ($10^{-2} \lesssim \rho_{14} \lesssim 1$).  Figure \ref{fig:BT} details the magnitude of the pulsar kick and the energy emitted in sterile neutrinos as a function of the magnetic field-temperature parameter space for these densities of interest in the protoneutron star.  In order for the active-sterile neutrino transformation mechanism to produce the pulsar kicks, for a given magnetic field strength, the temperature must be large enough to produce the fastest kicks ($\sim 1000~{\rm km}/{\rm s}$) yet small enough that the total energy emission in sterile neutrinos does not overcool the star ($\lesssim 10^{53}~{\rm erg}$).  

The active-sterile neutrino transformation mechanism for generating pulsar kicks requires large magnetic fields, $\sim 10^{16-17}~{\rm G}$, to generate the largest observed pulsar kicks.  These requisite magnetic fields need only be present for a fraction of a second in the lifetime of the collapsing protoneutron star for the mechanism to provide a sufficient pulsar kick.  The magnetic fields may be large, as the virial theorem allows for magnetic field strengths of $\sim 10^{18}~{\rm G}$ or more \cite{cf53,cbp97,pbc99}. Other considerations suggest that it is plausible that these large magnetic fields may occur inside some protoneutron stars for the timescales needed to generate the pulsar kick as described in this work \cite{dt92, kou99,pbc99,bah83}.

%However, in the interpretation of Figure \ref{fig:BT}, one must be aware of the caveats presented in Section \ref{sec:caveats} tend to suppress both the magnitude of the pulsar kick and the total energy emitted.

\section{Conclusions}
\label{sec:conclusions}

Ref.\ \cite{ks97} introduced the notion that if there was a sterile neutrino with a mass of order a few keV with a vacuum mixing with active neutrinos that pulsar kicks could be explained by resonant active-sterile neutrino transformations in a supernova explosion.  The authors argue that anisotropic emission of sterile neutrinos can create an asymmetry in the momentum emitted in sterile neutrinos which is a few percent of the total momentum emitted in sterile neutrinos.  The focus of this paper was to elaborate upon this mechanism toward providing pulsar kicks that can explain the large velocities of pulsars that have been observed.

In this work, a phase space-based approach was taken to describe the active-sterile neutrino transformation.  At any point in the star, a shell in phase space of active neutrinos have an MSW-like resonance and will transform into sterile neutrinos with an efficiency that is dictated by the adiabaticity of the transition.  If the width of the resonance is large compared to the neutrino oscillation length, the transition is adiabatic and active neutrinos are efficiently converted into sterile neutrinos.  On the other hand, as the width of the resonance gets smaller, the transition is less adiabatic and the transformation efficiency suffers.  The details of this transformation process depend on the density and temperature profiles within the neutron star and the magnitude and direction of the magnetic field.  

The result is that when the resonant conversion of active neutrinos to sterile neutrinos occur at a location within the star where the scattering rate is high, a large flux of sterile neutrinos can be created.  The form of the forward scattering potential, and thus the resonant momenta, creates an asymmetry in the emitted momentum of sterile neutrinos generated by the resonant active-sterile neutrino transformation.  For realistic values of density, temperature and magnetic field strength, an asymmetry on the order of tens of percent can be exhibited in the emitted sterile neutrinos.  With such a large asymmetry, enough sterile neutrinos may be emitted from the core of the supernova to provide the pulsar kick, yet not so many that the sterile neutrinos will overcool the star, contradicting the active neutrino observations from Supernova 1987A.  

However, there are a number of other considerations that are beyond the scope of this work, but are important to determine the magnitude of the pulsar kick and associated cooling by sterile neutrinos.  
%In all, processes such as the reabsorption of sterile neutrinos and the dynamical evolution of the protoneutron star can reduce the estimates of the pulsar kick and energy emitted by an order of magnitude or more.  
While the estimates presented in this work provide hopeful suggestions that the active-sterile neutrino transformation mechanism can produce the requisite pulsar kick without overcooling the star, it is necessary for these calculations to be made alongside supernova simulations to determine the strength of the pulsar kick and the magnitude of the cooling due to the emission of sterile neutrinos.

While the existence of a light sterile neutrino may continue to be an unresolved problem in particle physics, the possibility of a light sterile neutrino presents exciting opportunity in astrophysics to resolve unexplained phenomena such as dark matter and pulsar kicks.  Although further exploration is necessary to determine the magnitude of the effect, active-sterile neutrino transformation in a supernova presents an intriguing possible mechanism toward the explanation of such large velocities observed in pulsars.

\begin{acknowledgments}
I would like to thank Alex Kusenko, George Fuller, Alexander Kuznetsov and Nickolay Mikheev for useful discussions.  This work was supported by DOE grant DE-FG03-91ER40662 and NASA ATFP grant NNX08AL48G.
\end{acknowledgments}

\bibliography{nukick}

\end{document}